\documentclass[showpacs,preprintnumbers,amsmath,amssymb,floatfix]{revtex4}

\usepackage{graphicx}
\usepackage{dcolumn}
\usepackage{bm}

\usepackage{amsmath}
\usepackage{psfrag}
\usepackage{color}
\usepackage{latexsym,amsfonts}
\usepackage{graphicx}
\providecommand{\href}[2]{#2}   

\begin{document}

\def\conj#1{\stackrel{*}{#1}}
\def\la{\lambda}
\def\La{\Lambda}
\def\ve{\varepsilon}
\def\<{\langle}
\def\>{\rangle}
\def\half{{^1\!/_2}}
\def\n{h}
\def\GeV{{\rm\ GeV}}


\title{Target Normal Spin Asymmetry of the Elastic $ep$-Scattering at Resonance Energy}
\author{Dmitry~Borisyuk}
\author{Alexander~Kobushkin}%
\email{kobushkin@bitp.kiev.ua}
\affiliation{Bogolyubov Institute for Theoretical Physics\\
Metrologicheskaya str. 14-B, 03143, Kiev, Ukraine}

\date{\today}

\begin{abstract}
We study the target normal spin asymmetry for the reaction $ep \rightarrow ep$ at electron energy up to few GeV in the laboratory frame. The asymmetry is proportional to the imaginary part of the reaction scattering amplitude. To estimate the imaginary part of the amplitude we use the unitarity relation and saturate the intermediate hadron states by the proton and resonances from the first, second and third resonance regions. The resonance electromagnetic transition amplitudes, needed to evaluate the asymmetry, are taken from experiment.    
\end{abstract}

\maketitle

\section{Introduction\label{sec:Introduction}}
A study of the elastic $ep$-scattering is an important source of information about the internal structure of the proton. Due to the smallness of the fine structure constant, $\alpha\approx 1/137$, the first order perturbation term (the one-photon exchange) is assumed to give a main contribution to the electromagnetic transition amplitude. In the one-photon approximation the elastic $ep$-scattering is described by two quantities, the electric, $G_{E}(Q^2)$, and magnetic, $G_{M}(Q^2)$, form factors. 

The form factors $G_{E}$ and $G_{M}$ are usually extracted from cross section data by the Rosenbluth separation method. The database for  $G_{E}$ and $G_{M}$ obtained by this method shows that the ratio $G_{E}/G_{M}$ is approximately constant. Recently new precise measurements of the ratio $G_{E}/G_{M}$ were done at Jefferson Lab \cite{Jones,Gayou,Punjabi} by the recoil polarization method \cite{ARekalo1,ARekalo2} and yield significantly different results than the Rosenbluth separation method \cite{Arrington}. 
 
Because both measurements are based on the one-photon exchange approximation it is natural to assume that this discrepancy may be explained by the second order perturbation term in the $\alpha$ expansion. The $\alpha^2$ perturbation term should result in many effects. Its real part contributes to the cross-section and destroys the Rosenbluth formula. The imaginary part appears in one-particle polarization observables of the elastic $ep$-scattering, the target and beam spin asymmetry, which vanish in the one-photon approximation. The present experimental technique makes possible to measure such observables and thus to take under control effects beyond the one-photon approximation.

The aim of the present paper is a calculation of the target spin normal
asymmetry in the elastic $ep$-scattering at electron lab. energy
$E_{\rm lab} \lesssim 2 \GeV$.

The imaginary part of the scattering amplitude, which determines the asymmetry, is simply related (through the unitarity condition) to the electroproduction amplitudes of different hadronic states. The so-called ``elastic'' contribution (i.e. when the hadronic intermediate state, entering the unitarity condition, is the proton) to the asymmetry was calculated in \cite{Hey}. In \cite{deRuj} authors got strict bounds on the ``inelastic'' part of the asymmetry using the Schwartz inequality. But, as the authors noted themselves, those bounds highly overestimate the actual values of the asymmetry, especially at high scattering angles. 

In the recent work \cite{Pasq} the asymmetry was calculated with $N$ and $\pi N$ intermediate states. Such approach gives a reasonable approximation at low electron energies, but becomes worse as the energy increases.

Contrary to \cite{Pasq}, we will calculate the contribution of the resonances
in the intermediate state, namely, $P_{33}(1232)$, $D_{13}(1520)$, $S_{11}(1535)$,
$F_{15}(1680)$ and $P_{11}(1440)$, using their experimental electroproduction
amplitudes. Such approach may be justified at intermediate energy,
$E_\mathrm{lab}\lesssim$2~GeV. For the large momentum transfer region
parton model calculations would be more adequate \cite{Chen}. 

It was noted in \cite{Pasq}, that the single-spin asymmetry is sensitive
to the electroproduction amplitudes in a wide range of photon virtualities,
and this may be a new way to access resonance transition form factors.
The results of the present work may be useful in planning of such experiments.

The paper in organized as follows. In Section~\ref{sec:General} we derive general formalism for the asymmetry, in Sections~\ref{sec:Res}~and~\ref{sec:Fit} we explain how we describe the resonances and fit their electromagnetic transition amplitudes. Numerical results and conclusions are given Section~\ref{sec:Results}.
\section{General formulae \label{sec:General}}

\subsection{Notation}

We denote the initial electron and proton momenta $k$ and $P$, respectively,
and the final momenta $k'$ and $P'$. The transferred momentum is $q=k-k'$ ($q^2<0$),
and the c.m. energy squared is $s=(P+k)^2=(P'+k')^2$.
Time and space components of 4-momenta are denoted like $P = (\epsilon_P,\vec P)$.
$M$ is the proton mass, the electron mass is neglected.
We denote Dirac matrices $\gamma_\mu$ and use the short-hand notation
$\hat a$ for $a_\mu\gamma^\mu$.

Proton spinors with definite helicity $\la$ and momentum $P$ are
\begin{equation} \label{spinors}
  u_\la(P) = \left(
  \begin{array}{c}
    \sqrt{\epsilon_P + M}\ w_\la \\
    \sqrt{\epsilon_P - M}\ (\vec n \vec \sigma)\ w_\la
  \end{array} \right) {\rm ,\ where\ } w_\la = \left(
  \begin{array}{c}
    e^{-i \varphi/2} \cos {\theta + \pi(\half - \la) \over 2} \\
    e^{i \varphi/2} \sin {\theta + \pi(\half - \la) \over 2} \\
   \end{array} \right),
\end{equation}
where $\theta,\varphi$ are spherical angles of the vector
$\vec n = \vec P/|\vec P|$, $\vec \sigma$ are Pauli matrices.

Electromagnetic current matrix elements for the proton read
\begin{equation} \label{Jproton}
 \<P' \la' |J^\mu|P \la \> = \bar u_{\la'}(P')\ \Gamma^\mu\ u_\la(P) = 
 \bar u_{\la'}(P')
 \left[ 2M(G_E-G_M){P^\mu_+ \over P_+^2} + G_M \gamma^\mu \right] u_\la(P),
\end{equation}
where $|P\la\>$ is the proton state with momentum $P$ and helicity $\la$,
$P_+ = P+P'$, $G_E \equiv G_E(q^2)$ and $G_M \equiv G_M(q^2)$
are the proton elastic form factors.

\begin{figure}
  \hspace{-2.cm}
  \includegraphics[height=0.15\textheight]{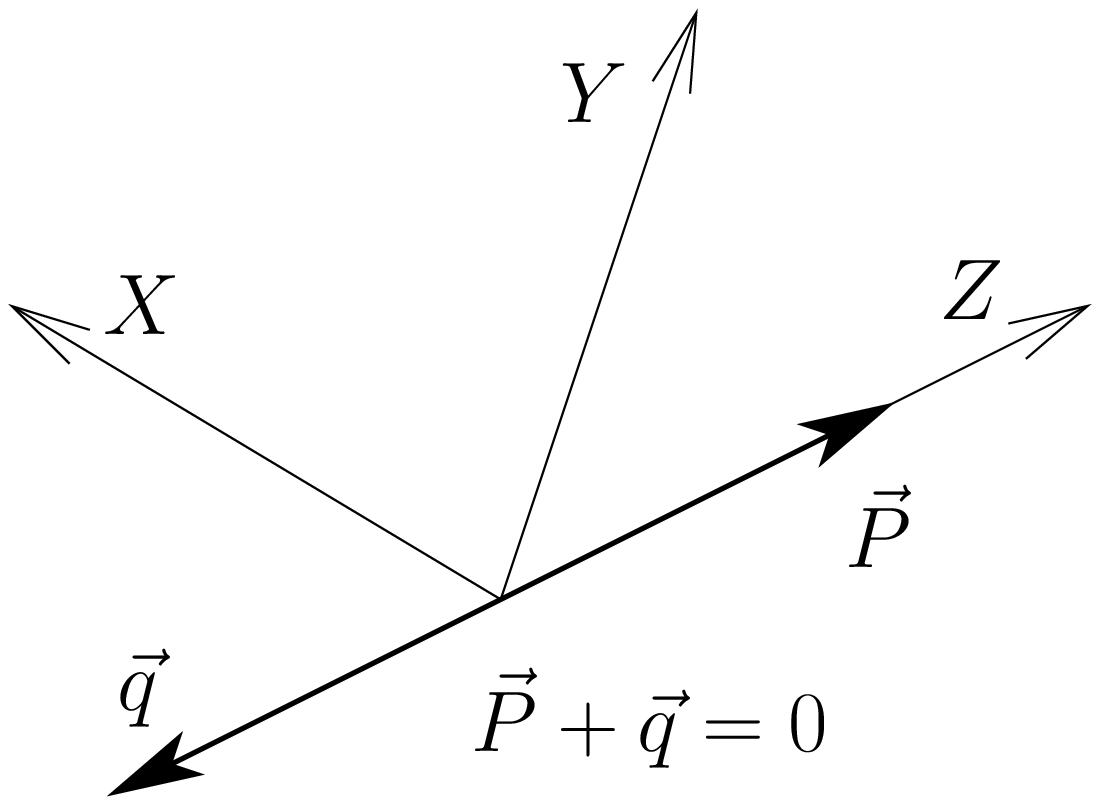}
  \hspace{4.cm}
  \includegraphics[height=0.2\textheight]{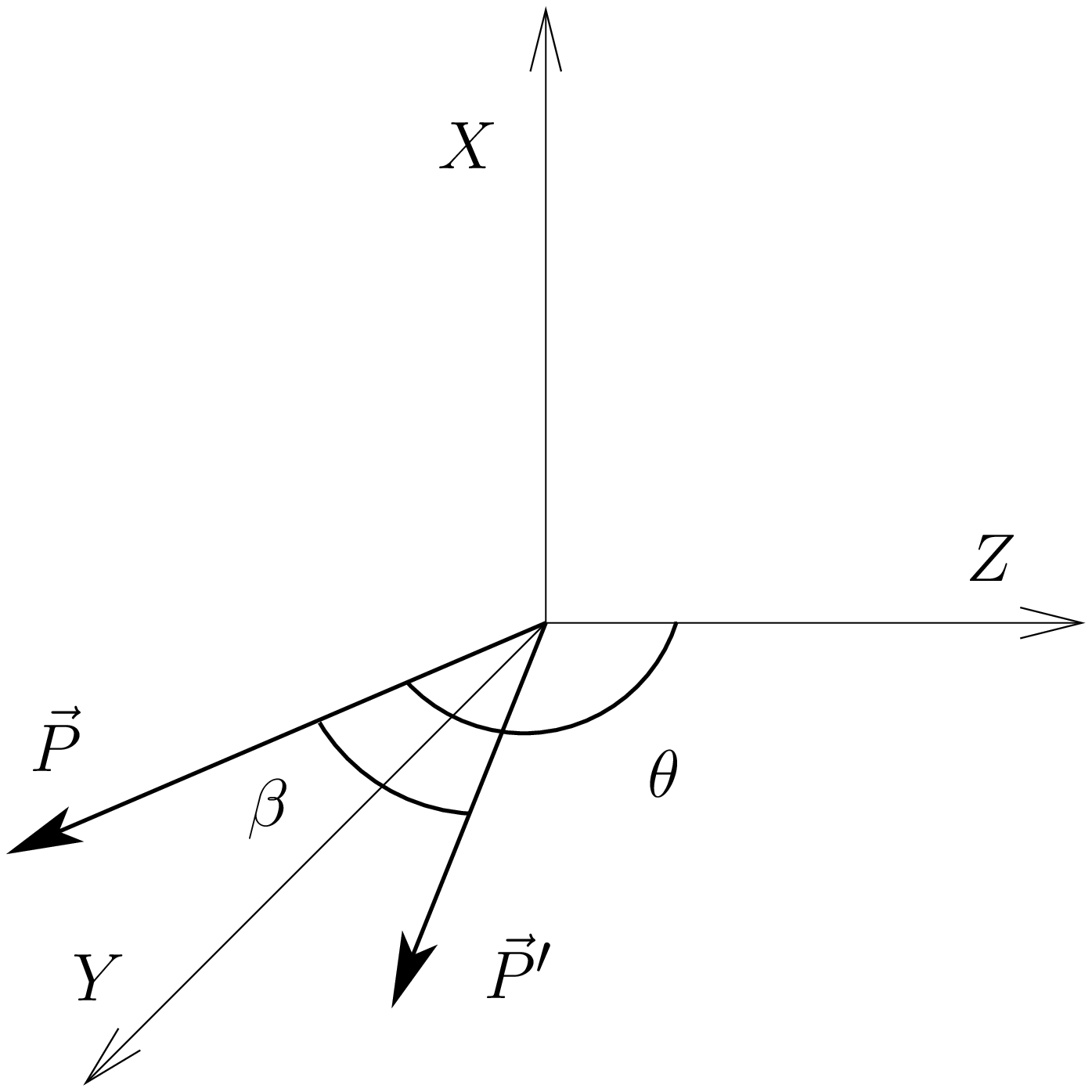}
  \\
  \parbox{0.47\textwidth}{\caption{\label{figefs} To the definition of transition amplitudes.}} 
   \parbox{0.47\textwidth}{\caption{\label{figbeta} To the derivation of the hadronic tensor.}}  
\end{figure}

Current matrix elements between the proton (with momentum $P$) and other hadronic states (with momentum $P''=P+q$) can be expressed via 3 independent invariant amplitudes. In the rest frame of the hadronic state (see Figure~\ref{figefs}) one has
\begin{equation} \label{efs}
  \ve^{(\la)}_\mu \<\n \La |J^\mu| P \half \> =
  f^{(\n)}_\la(q^2) \delta_{\La,\la + \half} ,\ \ \ \ 
  \ve^{(-\la)}_\mu \<\n -\!\La |J^\mu| P -\!\half \> =
  \eta_\n f^{(\n)}_\la(q^2) \delta_{\La,\la + \half}.  
\end{equation}
Here $\eta_\n = \pi_\n e^{i \pi(s_\n-\half)}$, $\n$ - some hadronic state,
$s_\n,\pi_\n$ - its spin and parity, $\La$ - spin projection onto the
vector $\vec P$. The quantities $f_\la^{(\n)}$ can be considered
as helicity amplitudes of the process $\gamma^\ast p \rightarrow \n$.

Polarization vectors of a virtual (space-like) photon $\ve_\mu$ are defined
according to \cite{Landau}. In the coordinate frame of Figure~\ref{figefs}
they are
\begin{equation}
  \ve^{(0)}_\mu = {1 \over \sqrt{-q^2}}(|\vec q\,|,0,0,-q^0), \qquad
  \ve^{(\pm 1)}_\mu = {1 \over \sqrt{2}}(0,\mp 1,-i,0),
\end{equation}
where the upper index of $\ve$ shows the spin projection onto the $z-$axis.

If the coordinate system is oriented arbitrarily, so that $\vec n = - \vec q/|\vec q\,|
= (\cos\varphi \sin\theta,\ \sin\varphi \sin\theta,\ \cos\theta)$, then
\begin{eqnarray} \label{polar}
  & \ve^{(0)}_\mu = {1\over\sqrt{-q^2}}(|\vec q\,|,\ -q^0 \vec n), \\
  & \ve^{(\pm 1)}_\mu = {1 \over \sqrt{2}}\left\{ (0,i \sin\varphi,\ -i \cos\varphi,\ 0)
  \mp (0,\cos\varphi \cos\theta,\ \sin\varphi \cos\theta,\ -\sin\theta)\right\}. \nonumber
\end{eqnarray}

Orthogonality relations
\begin{equation}
  \sum_\la (-1)^\la \ve^{(\la)}_\mu \conj{\ve}\!^{(\la)}_\nu
   = g_{\mu\nu} - {q_\mu q_\nu \over q^2} \quad \text{and} \quad 
  g^{\mu\nu} \ve^{(\la)}_\mu \conj{\ve}\!^{(\la')}_\nu =
  (-1)^\la \delta_{\la \la'}.
\end{equation}

\subsection{Asymmetry}

The name ``target normal asymmetry'' corresponds to the situation when
the {\it target} proton is polarized along the {\it normal} to the
reaction plane, and other particles are unpolarized. Under such condition
the proton spin has two possible directions, say, above and below the
reaction plane. Its invariant spin 4-vector should be either collinear
or anti-collinear with 4-vector
\begin{equation}
  S^\mu = {2 \ve^{\nu\mu\sigma\tau} k_\nu P_\sigma P'_\tau \over
             \sqrt{q^2(sq^2-(s-M^2)^2)}}\ ,
\end{equation}
which is orthogonal to all momenta and satisfies $S^2=-1$.
The corresponding cross-sections, $\sigma_\uparrow$
and $\sigma_\downarrow$, are equal in the one-photon approximation,
so the difference between them is due to higher order perturbative terms.
The target normal asymmetry is defined as dimensionless ratio
\begin{equation}
A_n = \frac{\sigma_\uparrow - \sigma_\downarrow}
         {\sigma_\uparrow + \sigma_\downarrow}.
\end{equation}
The asymmetry is proportional to the imaginary part of the scattering amplitude.
The imaginary part, in turn, can be expressed through the unitarity condition,
which reads
\begin{equation} \label{unit}
i(T_{fi}-\conj T_{if}) = \sum_n T_{fn} \conj T_{in},
\end{equation}
where $i$ and $f$ are initial and final states, respectively, $n$ is so-called
intermediate state and $T_{fi}$ are $T$-matrix elements.
In our case we can, as the first approximation, use one-photon exchange
amplitudes in the right-hand side of (\ref{unit}). Then we obtain
\begin{equation} \label{unit2}
\begin{split} \includegraphics[height=0.1\textheight]{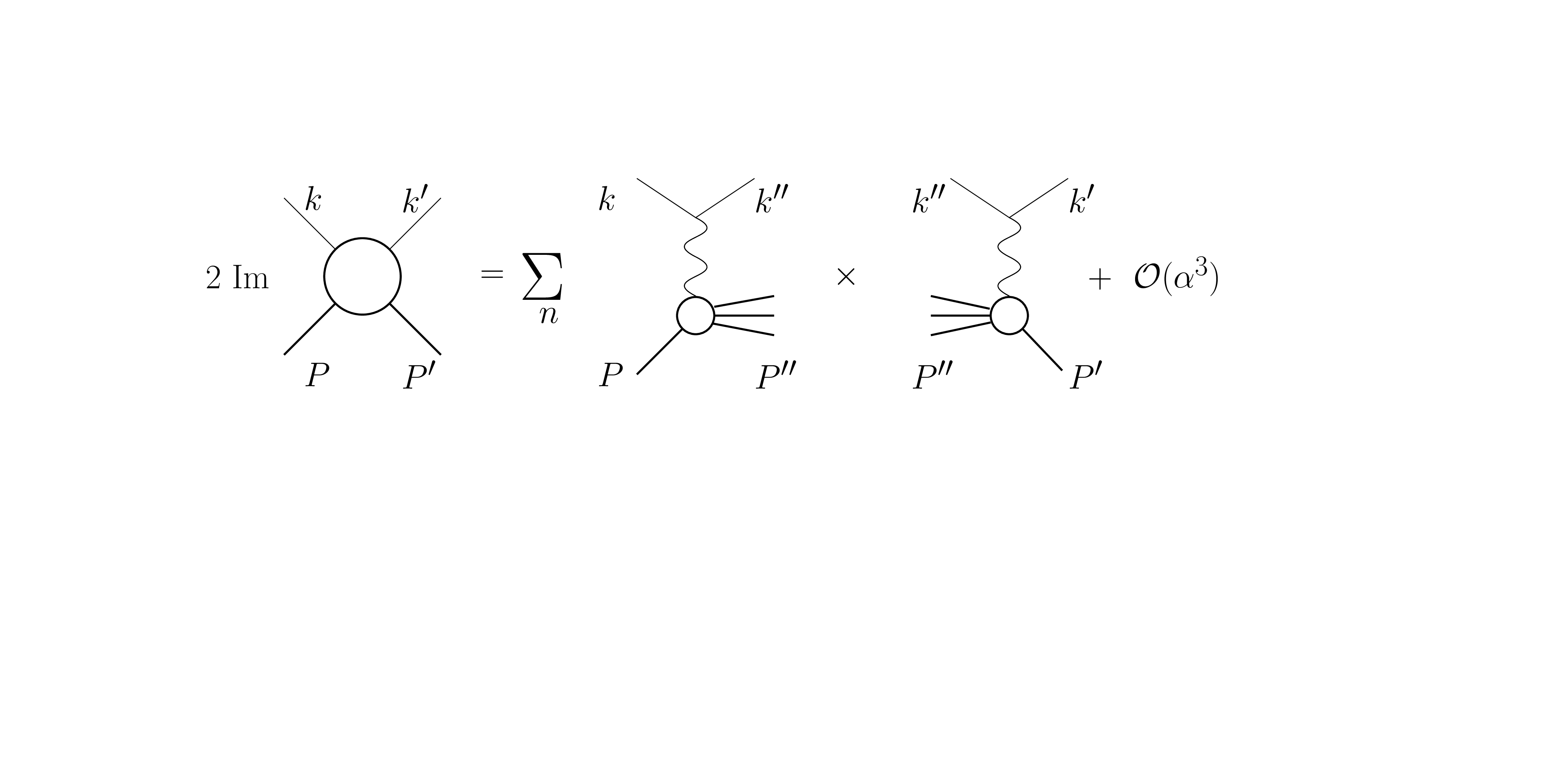} \end{split}
\end{equation}
We denote $q_1 = k-k''$, $q_2 = k'-k''$  and 
the mass of hadronic intermediate state $W=\sqrt{P''^2}$.

As it was shown in \cite{deRuj} using (\ref{unit2}) and the time-reversal symmetry
of the electromagnetic interaction, the asymmetry can be written as
\footnote{There is a misprint in Ref.~\cite{deRuj}, a missing factor 8 in the denominator.}
\begin{eqnarray} \label{asy2}
  A_n = {i \alpha q^2 \over 2 \pi^2 D}
    \int\limits_{M^2}^s {s-W^2 \over 8s} dW^2 \int d\Omega_{k''}
    {1 \over q_1^2 q_2^2} L^{\alpha\mu\nu} \sum_{\la_p,\la'_p}
    W_{\mu\nu}(P'\la'_p;P\la_p)\,
    \bar u_{\la_p}(P) (-\gamma^5 \hat S \Gamma_\alpha) u_{\la'_p}(P')  
\end{eqnarray}
in the first non-vanishing order of $\alpha$, where $\Gamma_\alpha$ is defined
in (\ref{Jproton}),
\begin{equation}
  D = 4 \left( { (2s + q^2 - 2M^2)^2 \over 4 M^2 - q^2}
  (4 M^2 G_E^2 - q^2 G_M^2) + q^2 (4 M^2 G_E^2 + q^2 G_M^2) \right),
\end{equation}
\begin{equation}
  L^{\alpha\mu\nu} =
  {\rm Tr} (\hat k' \gamma^\mu \hat k'' \gamma^\nu \hat k \gamma^\alpha),
\end{equation}
and the hadronic tensor $W_{\mu\nu}$ is defined as
\begin{equation}
  W_{\mu\nu}(P'\la'_p;P\la_p) = \sum_\n (2\pi)^4 \delta(P+k-P''-k'')
  \<P'\la'_p|J_\mu|\n\> \<\n|J_\nu|P\la_p\>.
\end{equation}
Here $|\n\>$ are all possible hadronic states, which we will refer to as
``intermediate states''. They can be $N$, $\pi N$, $\pi\pi N$, $\eta N$ and so on.
$\sum\limits_\n$ is the short-hand notation for
\begin{equation}
  \sum_N \sum_{\rm spins} \int \prod_{a=1}^N {d^3 p_a \over (2\pi)^3 2 \epsilon_{p_a}},
\end{equation}
where $N$ is the total number of particles, $p_a$ are their momenta.

Now we will express the hadronic tensor $W_{\mu\nu}$ through the
electroproduction helicity amplitudes $f^{(\n)}_\la$.

Consider this tensor in the rest frame of the hadronic state $|\n\>$, i.e. at $\vec P''=0$.
We choose coordinate system so that both vectors $\vec P$ and $\vec P'$ lie in
the $yz$-plane (see Figure \ref{figbeta}), and the angle between them is
$\beta$, $0\le\beta\le\pi$.

The sum $\sum\limits_\n$ can be split into 2 parts: first, the sum over
total angular momentum (=spin) projections, second, the sum over all remaining
quantum numbers, which we denote ${\sum\limits_\n}'$.
The behaviour of the state $|\n\>$ with respect to spatial rotations is
completely described by its spin and spin projection, and does not
depend on any other quantum numbers. Thanks to this fact, the sum over
spin projections can be done explicitly:
\begin{eqnarray}
  & W_{\mu\nu}(P'\la'_p; P\la_p) = {\sum\limits_\n}' \sum\limits_{\La''}
  (2\pi)^4 \delta(P+q_1-P'') \<P' \la'_p|J_\mu|\n \La''(\vec e_z)\>
  \<\n \La''(\vec e_z)|J_\nu|P\la_p\> = \\
  & = {\sum\limits_\n}' (2\pi)^4 \delta(P+q_1-P'') \sum\limits_{\La,\La',\La''}
  \<P' \la'_p|J_\mu|n \La'(\vec P')\>
  \<\n \La'(\vec P')|\n \La''(\vec e_z)\> \<\n \La''(\vec e_z)|\n \La(\vec P)\>
  \<\n \La(\vec P)|J_\nu|P\la_p\>. \nonumber
\end{eqnarray}
Here $|\n\La(\vec a)\>$ denotes the state with the spin projection
onto vector $\vec a$ equal to $\La$.

Wave functions of these states are related via Wigner D-functions \cite{Landau}:
\begin{equation} \label{wf}
  \<\n \La''(\vec e_z)|\n \La(\vec P)\> =
  \conj {\cal D}\!^{(s_\n)}_{\La \La''}(\varphi, \theta, 0),
\end{equation}
where $\varphi$, $\theta$ are polar angles of the vector $\vec P$
and $s_\n$ is the spin of the state $|\n\>$.
Using (\ref{wf}) and properties of D-functions we have
\begin{equation}
  \sum_{\La''} \<\n \La'(\vec P')|\n \La''(\vec e_z)\>
  \<\n \La''(\vec e_z)|\n \La(\vec P)\> = 
  \sum_{\La''} {\cal D}^{(s_\n)}_{\La' \La''} ({\textstyle {\pi \over 2}},\theta-\beta,0)
  \conj {\cal D}\!^{(s_\n)}_{\La \La''} ({\textstyle {\pi \over 2}},\theta,0) =
  {\cal D}_{\La \La'}^{(s_\n)}(0,\beta,0).
\end{equation}
Using also amplitudes definition (\ref{efs}), we obtain
\begin{eqnarray}
  & W_{\mu\nu}(P' \la'_p; P \la_p) =
  \sum\limits_{\la,\la'} (-1)^{\la + \la'}
  \ve_{1\nu}^{(2\la_p \la)} \conj{\ve}\!_{2\mu}^{(2 \la'_p \la')} \times \\
  & \times {\sum\limits_\n}' (2\pi)^4 \delta(P+q_1-P'')
  f^{(\n)}_\la(q_1^2) \conj f\!^{(\n)}_{\la'}(q_2^2)
  \ \eta_\n^{\la_p- \la'_p}
  {\cal D}_{\la_p(2\la+1),\la'_p(2\la'+1)}^{(s_\n)}(0,\beta,0), \nonumber
\end{eqnarray}
where $\ve_1$ and $\ve_2$ are polarization vectors of the 1st ($q_1$) and
2nd ($q_2$) photons of Eq.(\ref{unit2}), defined according to (\ref{polar}).

After that the asymmetry becomes
\begin{eqnarray} \label{asy3}
  A_n = {\alpha q^2 \over \pi D} \int\limits^s_{M^2} {s-W^2 \over 8s} dW^2
  \int d\Omega_{k''} {1 \over q_1^2 q_2^2}
  \times \\ \times
  {\sum\limits_\n}' (2\pi)^3 \delta(P+k-P''-k'')
  \sum\limits_{\la,\la'} f^{(\n)}_\la(q_1^2) \conj f\!^{(\n)}_{\la'}(q_2^2)
  X^{(\n)}_{\la\la'}(W,q_1^2,q_2^2), \nonumber
\end{eqnarray}
where
\begin{eqnarray}
 X^{(\n)}_{\la\la'}(W,q_1^2,q_2^2) = i \sum_{\la_p,\la'_p}
 (-1)^{\la+\la'} \eta_\n^{\la_p-\la'_p}
 {\cal D}^{(s_\n)}_{\la_p(2\la+1),\la'_p(2\la'+1)}(0,\beta,0)
 \times \\ \times
 L^{\alpha\mu\nu} \ve_{1\nu}^{(2\la_p \la)}
 \conj{\ve}\!_{2\mu}^{(2\la_p'\la')}
 \bar u_{\la_p}(P) (-\gamma^5 \hat S \Gamma_\alpha) u_{\la'_p}(P'). \nonumber
\end{eqnarray}

The quantities $X^{(\n)}_{\la\la'}$ can be calculated explicitly (provided the
proton form factors are known), and the only unknown in  
(\ref{asy3}) are electromagnetic transition amplitudes $f^{(\n)}_\la$.

\section{The model for transition amplitudes \label{sec:Res}}

Obviously, it is practically impossible to take into account $all$ allowed
intermediate states. To proceed further, we need to restrict these states somehow.
The authors of \cite{Pasq}, for example, included only $N$ and $\pi N$ states.
Although below the $\pi\pi N$ threshold ($E_{\rm lab} \approx 0.3 \GeV$) such
approach gives an exact result, one can expect that as the energy
increases, this approximation becomes worse, since more intermediate states
(e.g. $\eta N$, $\pi\pi\pi N$) will be possible.

In the present paper we use another way to model the intermediate states.
We treat them as a number of resonances and neglect the non-resonant
continuum contribution.
At present we cannot estimate the non-resonant contribution well enough,
but we can give qualitative arguments that it is small.

At a glance one may conclude that the relative size of the non-resonant
contribution will be approximately the same as in inelastic cross-sections
or structure functions. But actually it is likely to be much smaller
for the following reason. Contrary to strictly positive quantities,
such as cross-sections, the asymmetry can have either sign. Thus the
contributions from different non-resonant states will mostly cancel each
other. This is similar to the fact that the average of many uncorrelated
random quantities has much smaller dispersion that each of them.

At $P' = P$ the hadronic tensor $W_{\mu\nu}(P'\la'_p;P \la_p)$, which was
introduced in previous section, turns into the hadronic tensor of
inelastic $ep$-scattering. It is natural to assume that the
qualitative properties of both tensors are similar, so we should first look
what resonances contribute to the inelastic $ep$-scattering.

There are three prominent resonant peaks in  the inelastic $ep$ cross-section:
the so-called 1-st, 2-nd and 3-rd resonance regions.
 
The first resonance peak is due to the $\Delta$-resonance ($P_{33}(1232)$), the
second peak consists of $D_{13}(1520)$ and $S_{11}(1535)$.
There are many resonances which contribute to the third resonance region, but
there are serious arguments (see, e.g., Ref.~\cite{Stoler}) that the dominant contribution comes from
$F_{15}(1680)$. Moreover, it is the only one, for which the transition amplitudes are known.
Although the Roper resonance $P_{11}(1440)$ does not contribute significantly
to the inelastic $ep$-scattering \cite{Stoler}, we also included it in our
calculations.

For the proton in the intermediate state one has
\begin{equation}
  {\sum_\n}' (2\pi)^3 \delta(P+q_1-P'') = \int
  {d^3 P'' \over 2\epsilon_{P''}} \delta(P+q_1-P'') =
  \delta(W^2- M^2),
\end{equation}
The resonance, however,
has a mass $M_R\ne M$ and some finite width $\Gamma_R$, so we ``spread'' $\delta$-function with the relativistic Breit-Wigner formula:
\begin{equation}
  \delta(W^2-M_R^2) \rightarrow {\Gamma_R M_R \over \pi}
  {1 \over (W^2-M_R^2)^2 + M_R^2 \Gamma_R^2}.
\end{equation}
After that the expression ${\sum\limits_\n}' (2\pi)^3 \delta(P+q-P'')
f^{(\n)}(q_1^2) \conj f\!^{(\n)}(q_2^2)$, entering the formula for asymmetry,
will take the form
\begin{equation}
  f^{(p)}(q_1^2) \conj f\!^{(p)}(q_2^2) \delta(W^2-M^2) + 
  \sum_R f^{(R)}(q_1^2) \conj f\!^{(R)}(q_2^2) {\Gamma_R M_R \over \pi}
  {1 \over (W^2-M_R^2)^2 + M_R^2 \Gamma_R^2}.
\end{equation}
The first part is the ``elastic'' (proton) contribution; in the second part
the sum runs over all resonances taken into account.
The quantities $f^{(R)}$ depend only on $q^2$ but not $W$. They are related
to commonly used \cite{exp} $A_{3/2}$, $A_{1/2}$, and $S_{1/2}$ as
\begin{equation}
f_1 = \kappa A_{3/2},\quad
  \eta_R f_{-1} = \kappa A_{1/2},\quad \text{and} \quad
 f_0 = {2 q^2 M_R \over \sqrt{4M^2q^4-q^2(M_R^2-M^2-q^2)^2}} 
  \kappa S_{1/2},
\end{equation}
where $\kappa = \sqrt{M(M_R^2-M^2) \over \pi\alpha}$, while for the proton they are
%
%
\begin{equation}\label{proton}
  f_1^{(p)}(q^2)   \equiv 0, \quad
  f_0^{(p)}(q^2)    =  2MG_E(q^2), \quad
  f_{-1}^{(p)}(q^2)  = -G_M(q^2) \sqrt{-2q^2}, 
\end{equation}
which is easy to derive comparing (\ref{Jproton}) and (\ref{efs}).

For our calculation we use experimental data on $A_H$ (i.e. $A_{3/2}$,
$A_{1/2}$, $S_{1/2}$) given in \cite{exp}. Unfortunately, there are no data
on $S_{1/2}$ for $D_{13}(1520)$ and $F_{15}(1680)$, so in our calculations we
set it to zero.

To evaluate the asymmetry we need to fit these data somehow. The fitting 
procedure is described in the next section.
\section{Fitting procedure \label{sec:Fit}}
Masses and widths of resonances were taken from PDG \cite{PDG}.
The proton form factors were modelled using the well-known dipole fit
\begin{equation}
  G_M(q^2)/\mu_p = G_E(q^2) = {1 \over (1+Q^2/Q_0^2)^2},
\end{equation}
where $Q^2 \equiv -q^2$, $Q_0^2=0.71 \GeV^2$ and $\mu_p\approx2.79$
is the proton magnetic moment.

Now consider the fit of electroproduction amplitudes.
According to quark model prediction the high $Q^2$ behaviour of the transition amplitudes should be like $A_{1/2} \sim S_{1/2} \sim Q^{-3},\ A_{3/2} \sim Q^{-5}$ \cite{CM}.

For the proton we have
\begin{equation} \label{egproton}
  A_{1/2} \sim \sqrt{-q^2} G_M(q^2) \sim
  {Q \over (1+Q^2/Q_0^2)^2} \sim Q^{-3}.
\end{equation}
The denominator (dipole formula) is entirely due to quark structure,
while the numerator is just a kinematical factor. But the fact that
$A_{1/2}$ tends to zero as $Q^2\to 0$ is the specific feature of the proton.
For the other hadronic states $A_{1/2}(0) \neq 0$, so like \cite{CM}
and for the same reasons we introduce factor $\sqrt{(M_R-M)^2 + Q^2}$
and assume, instead of (\ref{egproton}),
\begin{equation} \label{asymp}
 A_{1/2} \sim {\sqrt{(M_R-M)^2 + Q^2} \over (1+Q^2/Q_0^2)^2}.
\end{equation}
So, to obtain correct asymptotic behaviour of $A_{1/2}$ at $Q^2\rightarrow 0$ and
$Q^2\rightarrow \infty$, it is useful to fit the function
\begin{equation} \label{Atilde}
  \tilde A_{1/2} = { (1+Q^2/Q_0^2)^2 \over \sqrt{(M_R-M)^2 + Q^2}} A_{1/2},
\end{equation}
which has finite values at both $Q^2=0$ and $Q^2=\infty$.
The same can be stated for $S_{1/2}$. On the other hand, $A_{3/2}$ has another
asymptotic behaviour ($\sim Q^{-5}$), therefore we would have used
$(1+Q^2/Q_0^2)^3$ instead of $(1+Q^2/Q_0^2)^2$ in expressions like
(\ref{asymp}) and (\ref{Atilde}) for $A_{3/2}$. 

However the $Q^2$ values needed for our calculation are not too high.
Since $Q^2 = {(s-M^2)^2 \over s} \sin^2 {\theta\over 2}$,
where $\theta$ is the c.m. scattering angle, one gets $Q^2_\mathrm{max} \sim \ 3\ \mathrm{GeV}^2$ at $E_{\rm lab}=2\GeV$.
%
%
Trying different parameterizations, we found that better agreement with the experimental data
in the range $Q^2 \lesssim 3 \GeV^2$, especially for Delta, is achieved
if we use the same formulae
\begin{equation}
 A_{H} \sim {\sqrt{(M_R-M)^2 + Q^2} \over (1+Q^2/Q_0^2)^2}
\end{equation}
for all amplitudes. Thus we fit the functions
\begin{equation}
  \tilde A_{H} = { (1+Q^2/Q_0^2)^2 \over \sqrt{(M_R-M)^2 + Q^2}} A_{H}.
\end{equation}
To describe high $Q^2$ behaviour better, we treat $\tilde A_H$ as a function
of $\xi = 1- {1 \over 1+Q^2/Q_0^2}$ instead of $Q^2$. At low $Q^2$ it does not
matter, since $\xi \sim Q^2$, but the advantage is that $\xi$ is finite
at $Q^2\rightarrow\infty$, so we can use simple linear or polynomial fit for
all $\xi$ values.

\begin{figure}
  \includegraphics[height=0.22\textheight]{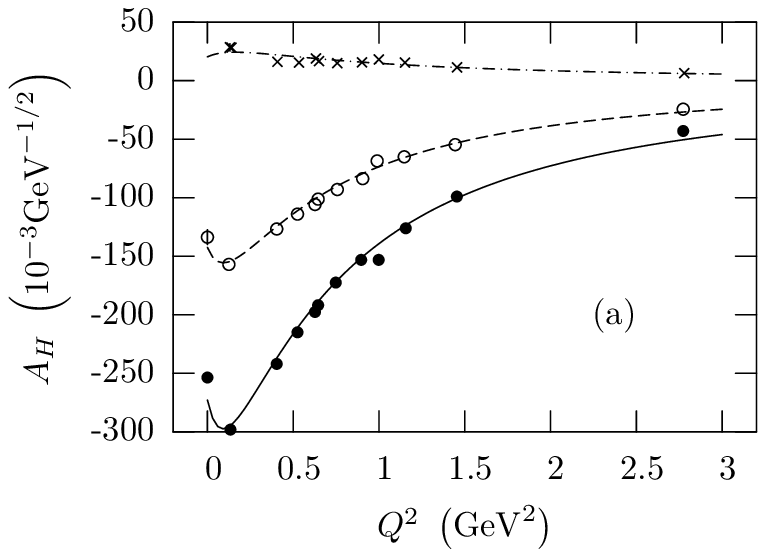}\\[0.5cm]
  \includegraphics[height=0.22\textheight]{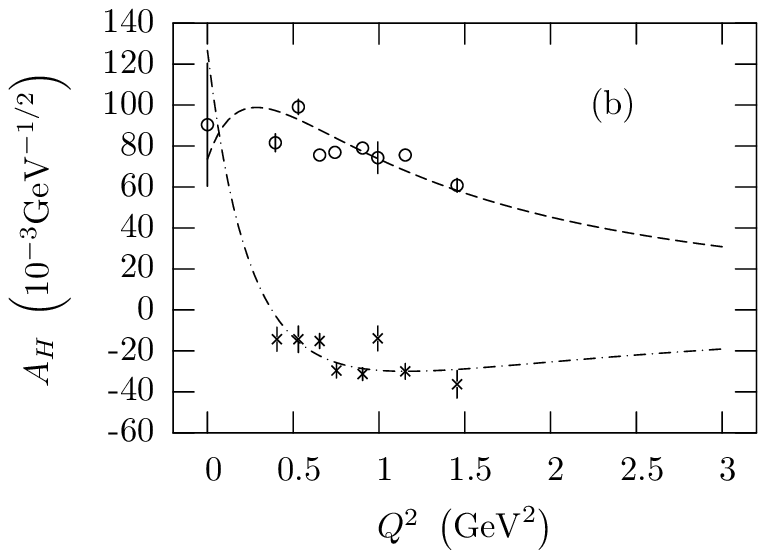}
  \includegraphics[height=0.22\textheight]{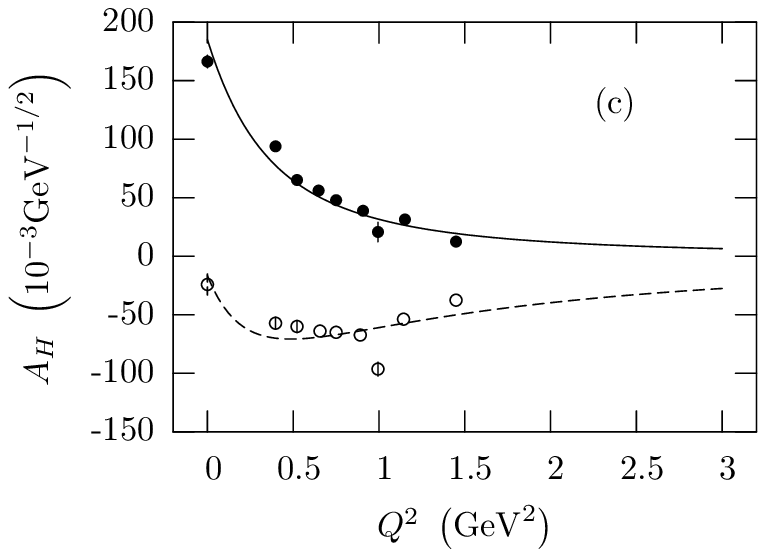}\\[0.5cm]
  \includegraphics[height=0.22\textheight]{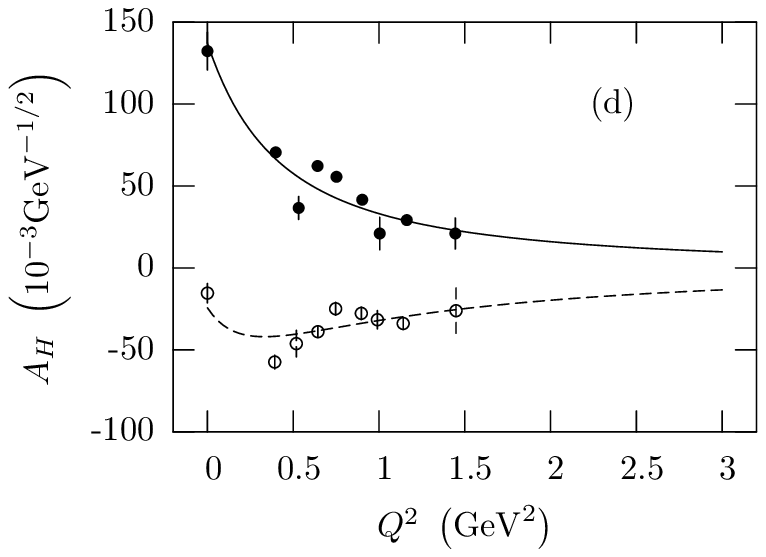}
  \includegraphics[height=0.22\textheight]{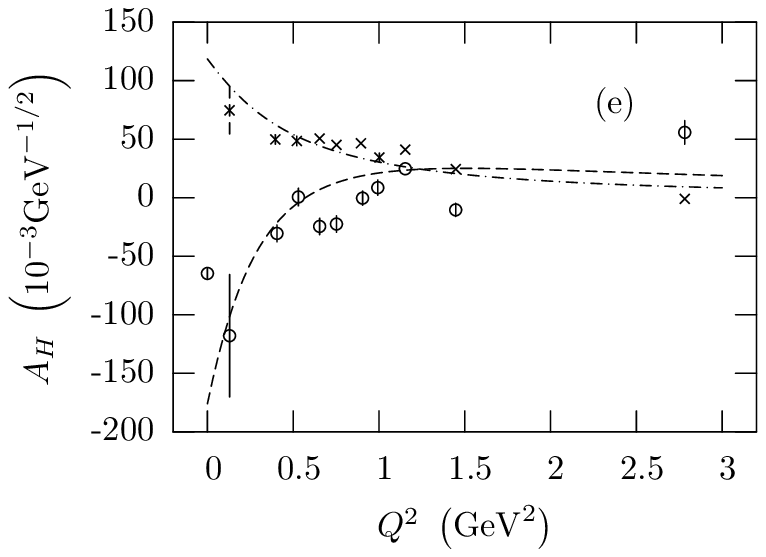}
\caption{\label{figfit}Fit of the transition amplitudes $A_{3/2}$
(solid lines and filled circles), $A_{1/2}$ (dashed lines and open
circles) and $S_{1/2}$ (dash-dotted lines and crosses)
for the resonances (a) $P_{33}(1232)$, (b) $S_{11}(1535)$, (c) $D_{13}(1520)$,
(d) $F_{15}(1680)$ and (e) $P_{11}(1440)$.
Experimental points are from a compilation of \cite{exp}.}
\end{figure}

In our calculations we restrict ourselves to the linear least-squares fit of the form
\begin{equation}
  \tilde A_H(\xi) = a + b\xi.
\end{equation}
Results of the fit are summarized in Table 1. The corresponding dependence of
amplitudes $A_H$ vs $Q^2$ for all considered resonances is shown
on Figure~\ref{figfit} together with experimental points.

\begin{table}
\centering
$\begin{array}{|c|r@{}lr@{}lr@{}l|}
\hline
\parbox{0.2\textwidth}{\centering \vspace{2mm} State \vspace{2mm}}
& & \tilde A_{3/2} & & \tilde A_{1/2} & & \tilde S_{1/2} \\
\hline
P_{33}(1232) & -0.929 & +0.264\,\xi  & -0.485 &+0.130\,\xi   & 0.069  &+0.022\,\xi \\
D_{13}(1520) & 0.318  & -0.273\,\xi & -0.029  &-0.474\,\xi  & & $no data$ \\
S_{11}(1535) & 0      &             & 0.123   &+0.416\,\xi  & 0.212 &-0.614\,\xi \\
F_{15}(1680) & 0.185  & -0.052\,\xi  & -0.033  & -0.199\,\xi & & $no data$ \\
P_{11}(1440) & 0      &             & -0.351 &+0.787\,\xi  & 0.236 &-0.134\,\xi \\
\hline
\end{array}$
\caption{Fit of the transition amplitudes.
All values are in $\GeV^{-1}.$}
\end{table}

\section{Numerical results and concluding remarks \label{sec:Results}}
\begin{figure}   
  \includegraphics[height=0.24\textheight]{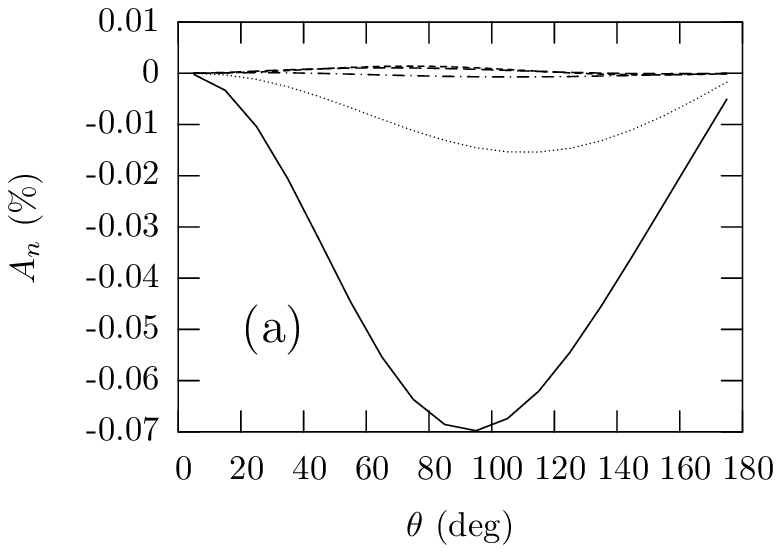}
  \includegraphics[height=0.24\textheight]{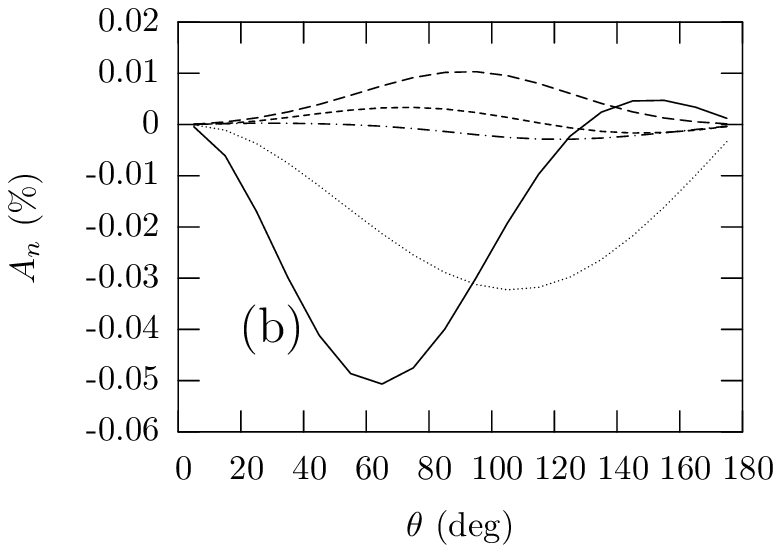}\\
  \includegraphics[height=0.24\textheight]{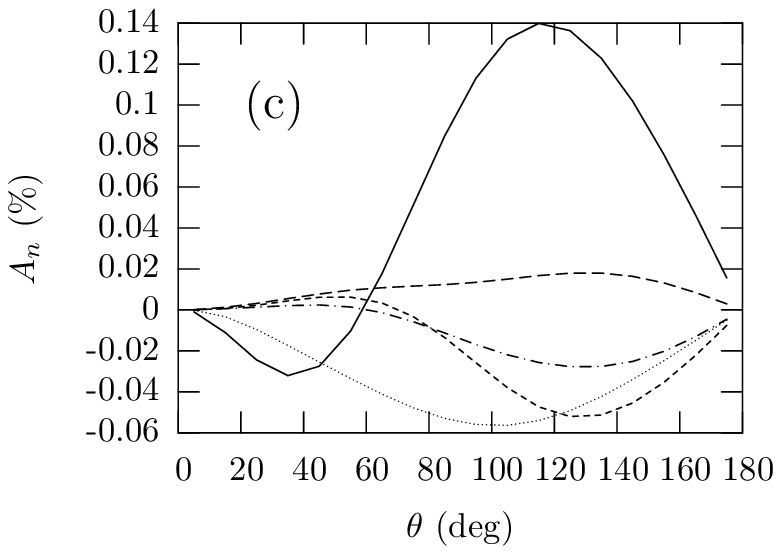}
  \includegraphics[height=0.24\textheight]{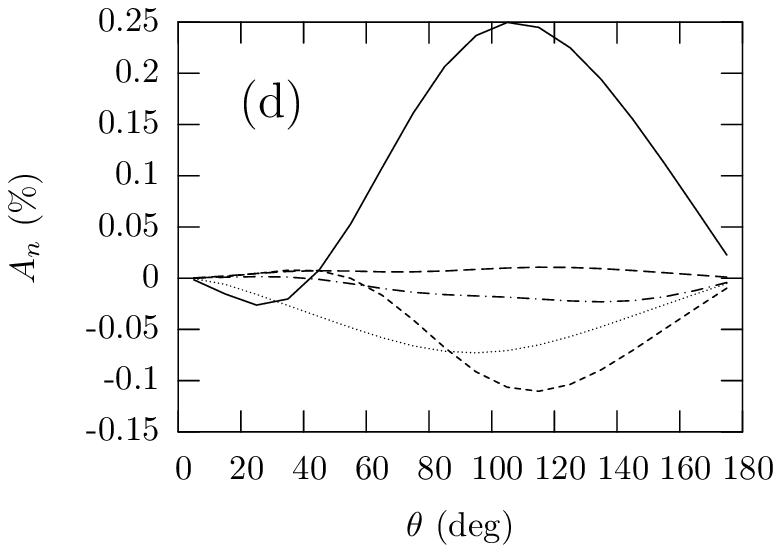}
\caption{\label{figinel} The contribution of resonances to the asymmetry at
different electron lab. energies, (a) 0.57 GeV, (b) 0.855 GeV, (c) 1.4 GeV
and (d) 2 GeV. Solid line --- $P_{33}(1232)$, long-dashed line ---
$D_{13}(1520)$, short-dashed line --- $S_{11}(1535)$, dash-dotted
line --- $F_{15}(1680)$ and dotted line --- $P_{11}(1440)$.}
  \includegraphics[height=0.24\textheight]{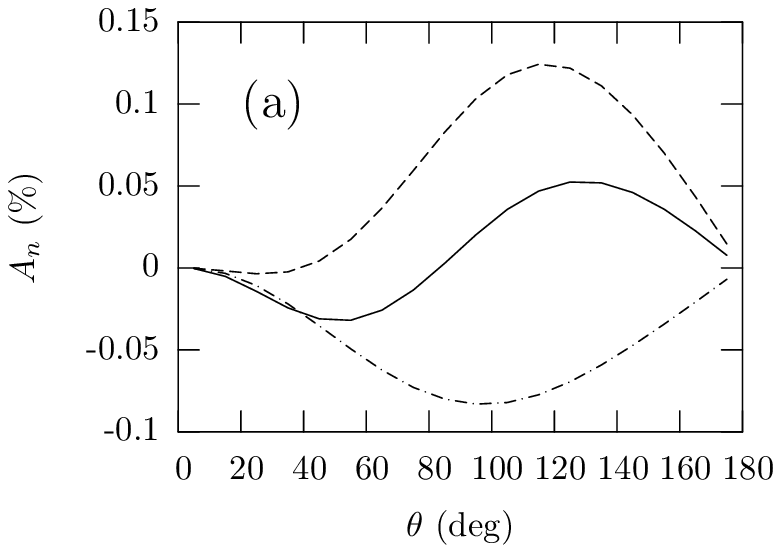}
  \includegraphics[height=0.24\textheight]{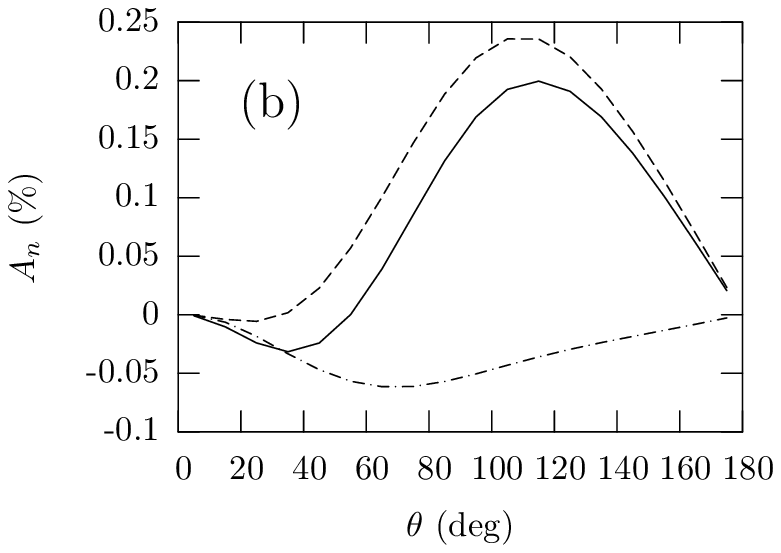}\\
  \includegraphics[height=0.24\textheight]{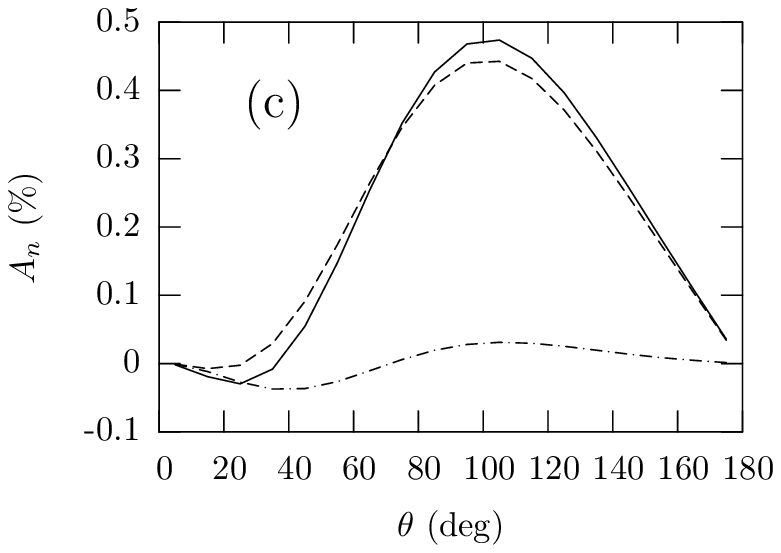}
  \includegraphics[height=0.24\textheight]{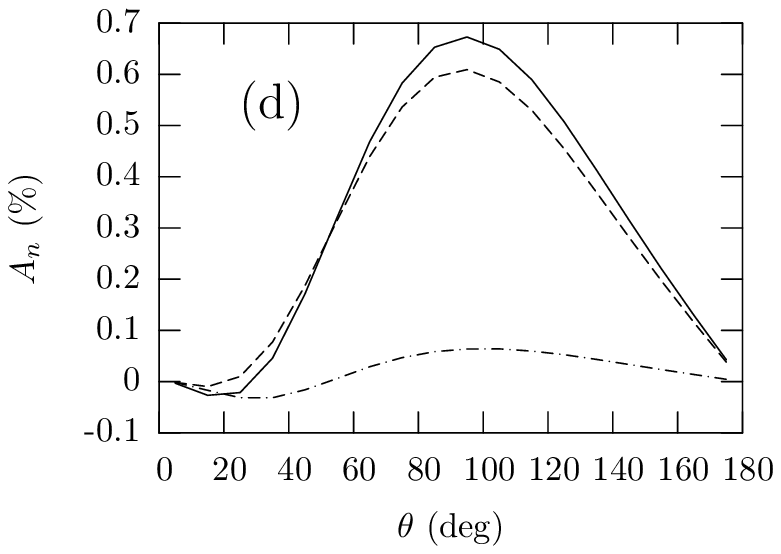}
\caption{\label{figtot} Target normal spin asymmetry for different electron lab.
energies, (a) 0.57 GeV, (b) 0.855 GeV, (c) 1.4 GeV and (d) 2 GeV.
The dashed line is the elastic contribution, the dash-dotted line is the
inelastic contribution, the solid line is total.}
\end{figure}

Figure~\ref{figinel} displays the contribution of separate resonances
to the target normal asymmetry $A_n$ vs the c.m. scattering angle $\theta$
at different electron lab. energy, $E_\mathrm{lab}$. One sees that globally
the $\Delta(1232)$ contribution is dominant. This is due to its large
transition amplitudes, in comparison with other resonances, and the lowest
mass among them. The contribution of the Roper resonance was obtained to be
not negligible. Moreover at $E_\mathrm{lab}\sim$0.9 GeV it becomes comparable
with the $\Delta(1232)$ contribution (the upper-right panel of
Figure~\ref{figinel}). This is very nontrivial fact, because the Roper
contribution in inelastic $eN$-scattering is very small. Nevertheless it can
be studied in precise measurements of the $A_n$ at special kinematical conditions.

One sees also that the contributions 
from the $\Delta$ and other resonances have mostly opposite sign and tend to cancel each other, 
especially at high beam energy. It is clearly seen
from Figure~\ref{figtot}, where we plot the elastic (proton) and inelastic (resonance)
parts of the asymmetry and the total asymmetry. The elastic contribution dominates at low 
energy ($E_\mathrm{lab}<$0.3 GeV) and at energy higher than 1.3 GeV. 
It is quite obvious for the low energy, because
the energy is insufficient for resonances to be produced. But at high energy
it is nontrivial result, which has interesting consequences.
As was discussed in Introduction, the asymmetry depends on the imaginary part
of the amplitude. But since the real and imaginary parts are connected
(via the dispersion relations), we may expect that the {\it real} part will also
be defined mostly by proton contribution. This is important for the proper
interpretation of the proton form factors measurements.


In summary we have calculated the target normal spin asymmetry, $A_n$, for the
$e^-p\to e^-p$ reaction at the electron beam energy up to few GeV in the laboratory frame. This quantity gives a direct information about the imaginary
part of the reaction scattering amplitude and comes from the second and higher
order perturbative terms.

To calculate the imaginary part of the amplitude we used unitarity and saturated
the intermediate  hadron states by the proton (the so-called elastic contribution)
and the resonances from the first, second and third resonance regions
(the inelastic contribution).
We neglect the non-resonant inelastic contribution, which we expect to be small
(see section \ref{sec:Res}). Besides that, the calculated contributions
of separate resonances are interesting alone.

Our calculations demonstrate that at special kinematical conditions (the electron lab. energy near 0.9~GeV) the contribution of the Roper resonance $P_{11}(1440)$ becomes comparable with the $\Delta(1232)$-contribution and affects significantly the target asymmetry. It turn, this opens a possibility to study $P_{11}(1440)$ electromagnetic transition amplitudes in precise measurement of the asymmetry.

\def\Jou#1#2#3#4#5{#1, #2 {\bf #3}, #5 (#4)}

\end{document}